# Ten simple rules for measuring the impact of workshops

Shoaib Sufi[1], Beth Duckles[2], Iveta Simera[3], Terhi Nurmikko-Fuller[4], Louisa Bellis[5], Wadud Miah[6], Adriana Wilde[7], Aleksandra Nenadic[1], Raniere Silva[1], Jennifer A. de Beyer[3], Caroline Struthers[3], Iain Emsley[8], Olivier Philippe[9], Melissa Balzano[10], Sara Coelho[11], Heather Ford[12], Catherine Jones[13], Vanessa Higgins[14]

[1] School of Computer Science, University of Manchester, Manchester, United Kingdom

[2] Portland State University, Portland, Oregon, United States of America

[3] UK EQUATOR Centre, Centre for Statistics in Medicine, NDORMS, University of Oxford, Oxford, United Kingdom

[4] College of Arts and Social Sciences, Australian National University, Canberra, Australia

[5] Cancer Research UK Cambridge Centre, University of Cambridge, Cambridge, United Kingdom

[6] Numerical Algorithms Group, Oxford, United Kingdom

[7] School of Computer Science, University of St Andrews, St Andrews, Scotland, United Kingdom

[8] Oxford e-Research Centre, University of Oxford, Oxford, United Kingdom

[9] Electronics and Computer Science, University of Southampton, Southampton, United Kingdom

[10] ELIXIR Hub, Wellcome Genome Campus, Cambridge, United Kingdom

[11] EGI Foundation, Amsterdam, The Netherlands

[12] School of Media and Communication, University of Leeds, Leeds, United Kingdom

[13] Scientific Computing Department, Science and Technology Facilities Council, Rutherford Appleton Laboratory, Didcot, United Kingdom

[14] School of Social Sciences, University of Manchester, Manchester, United Kingdom

* Corresponding author

E-mail: shoaib.sufi@manchester.ac.uk

# Abstract

Workshops are used to explore a specific topic, transfer knowledge, solve identified problems or create something new. In funded research projects and other research endeavours, workshops are the mechanism to gather the wider project, community or interested people together around a particular topic. However, natural questions arise: how do we measure the impact of these workshops? Do we know whether they are meeting the goals and objectives we set for them? What indicators should we use? In response to these questions, this paper will outline rules that will improve the measurement of the impact of workshops.

## Author summary

The idea for this paper came from a workshop entitled ‘Measuring the Impact of Workshops’ (1). ‘Measuring the Impact of Workshops’ collected practices and ways of thinking from a diverse set of experienced workshop organisers. This paper summarises these ideas into a coherent set of recommendations, Ten Simple Rules, that should make measuring impact more straightforward and more intentional.

# Introduction

Why should we measure the impact of the workshops we organise and run? With good measurements, we can convince funders to maintain and support the work that we do, encourage people to attend and feel satisfied that the work that we are doing with our workshops is worthwhile and making a positive difference.

A consistent approach to measuring similar workshops allows us to compare them over time and show improvement or the need for adapting the workshop to be more successful for the intended audience.

Effective measurement is the precursor to evaluating the workshops that we organise, which allows us to make quality assertions; i.e., that our workshops deliver the benefits to stakeholders (funders, attendees, and ourselves) that we think they do. Workshops to provide training or information dissemination are a recognised communication and engagement activity that funders class as pathways to impact (2) (3). Impact is becoming increasingly important for assessing research, for example, the UK Research Excellence Framework (REF) has increased its weighting for impact from research from 20% to 25% for REF 2021 (4).

This paper focuses on measuring impact and the thinking, knowledge, skills and techniques surrounding this. There are other excellent resources for organising (5) (6), curating, facilitating (7), and improving interactivity (8) at workshops, meetings and unconferences (9). The reader is encouraged to consult them for broader information about effective workshop organisation and running.

This paper proposes 10 simple rules for measuring the impact of workshops. Rules 1 and 2 concern planning: what you need to think about to set the right context for being able to measure impact. Rules 3-5 are knowledge and skills-based rules: things to be aware of or know how to do before constructing the method of measuring impact. Rules 6-10 are techniques that can be used to improve how we measure workshop impact.

## Types of workshops

This guide is focused on three types of workshops, which are explained below for reference and to provide a consistent terminology. Rules 1-6, 8 and 10 apply to all three workshop types. Rules 7 (to understand a change in skills) and 9 (to assess skills learnt) are of particular interest and use for learning workshops although they could also be used for the other workshop types. We will also illustrate, where necessary, which rules are of particular use for specific workshop types.

### Exploratory workshops

In exploratory workshops, ideas are analysed to better understand a topic and its associated problems, current solutions and future challenges. These workshops can have aims such as identifying what actions are needed to move a particular topic forward or getting expert advice from and into different communities. The keynotes, lightning talks, mini-workshops, and discussion sessions at the Collaborations Workshop (10) series are an example of exploratory workshop sessions.

### Learning workshops

In learning workshops, a particular skill set, application, or technique is taught. The expected outcome is increased knowledge, competence, or confidence in a particular area or set of techniques. Examples of learning workshops are the Software Carpentry (11) and Data Carpentry (12) workshops. Such workshops typically include practical exercises to apply the knowledge gained with assistance provided by the workshop organisers.

### Creating workshops

Creating workshops bring together individuals with a common or intersecting interest to solve particular problems by collectively building something. They can include multidisciplinary teams where problem holders guide the creative process. What is made can vary; it could be software, standards, resources or even papers. Workshops in the humanities, where collections of researchers work on the translation or annotation of historical texts, are more akin to creating workshops than traditional exploratory or learning workshops. The commonly termed “hackathons” (13) are considered creating workshops for the purpose of this paper.

## Rule 1: Setting goals effectively

When developing your workshop, the workshop goals or objectives (14) need to be set. Define the outputs (what you want to produce) and outcomes (what difference you hope it will make) (15) for the workshop.

Although goal-setting is, in general, good workshop organisational practice, we highlight it here as it is the foundation on which impact can be measured, as we assess to what extent we met our stated goals.

One way of effectively measuring impact is to include those who will attend the workshop when creating the ultimate goals. By using the answers from a pre-workshop questionnaire, the attendees can influence the creation of the goals. Their answers can help set goals such as the change in skills needed, topics that should be explored and problems that need addressing. This can then guide what questions will be asked post-workshop in order to measure the overall impact.

Collecting feedback during workshop development could lead to small, but significant, adjustments in the programme to help meet workshop goals.

## Rule 2: Balancing time, effort and costs

It is important to take into account the cost of a workshop, in terms of time, effort, and money, when thinking about measuring its impact. The number of people, the duration, the venue (whether held in person or online), the price to the individual, the resources available within the organisation, and whether the workshop is one of a series or a stand-alone event can all affect how much effort is reasonable to put into measuring impact and which of the rules, below, are applied. For example, the impact of a free one-hour online workshop might be adequately analysed using a few survey-type questions sent out at the end of the workshop. However, a multi-day, moderately expensive workshop that uses a mixture of learning and exploring and intends to enthuse people about changing practice may require more effort to be put into the impact analysis, so more of the rules would come into play (especially Rules 6-10, related to techniques).

It is expensive to fully measure longer-term impact, such as how people are applying what they have learnt, or to establish a causal link between workshop attendance and improved research (see Behaviour and Results in the Kirkpatrick model (16)) (17). However, it is better to do at least some work in this space, imperfect as it may be (Rule 8), rather than insisting on measuring things perfectly or not at all (18).

## Rule 3: Create metrics purposefully

The process of taking a concept and converting it into a metric is called commensuration. Any time that we quantify something that is not easily turned into a metric, such as an idea like “satisfaction” or “comfort,” we are engaging in commensuration (19). Examples of commensuration include creating workshop evaluations, measures of job productivity such as human resource documents, and cost/benefit analyses.

Metrics help to make things more comparable, simplify complex information, and create standards that support easier decision-making. However, we should be aware of the context and assumptions made when a specific metric is created.

The metrics we form are ultimately made up of what we believe is important; we are a part of what we are trying to study. Therefore, metric formation (i.e., commensuration) has inherent bias. We often find ourselves measuring that which is easy to measure or that we most want metrics on. Being aware of this limitation allows us to be more honest and intentional about trying to minimise bias. By knowing what we value and what is easy to measure, we can examine our analysis and check where we are missing data.

When measuring the outcome of workshops, it is important to ask questions that will elicit useful responses to help us answer our research questions or goals and support our intended analysis of the results of the question. This does not mean that we should bias the research towards a particular end. Instead, we should gather data that is useful for the task of discovering the concepts and outcomes that matter most to us, whether through scoring, categorisation, or free-text responses.

# Rule 4: Understand bias

The work of controlling biases is never finished. An iconic study in the field of management sciences, the Hawthorne effect (20), showed that the act of studying other humans will affect the outcome of the study. We can only evaluate our data honestly if we know what our biases are and are willing to be open about where they might be coming from.

## Common biases

Table 1 gives a non-exhaustive list of the common biases that can affect impact measurement (21):

*Table 1: Common biases and countermeasures*

| Bias type | Explanation | How to counter |
|---|---|---|
| **Confirmation bias** | The tendency to reaffirm your own values and beliefs, and to create research methods that confirm what you already believe to be true. For instance, I might decide that I'd like evidence that my workshops are | Know what we believe to be true and make certain that the questions allow for the opposite (and other) responses. |
| | very effective, so I ask questions designed to get mostly positive responses. | |
| **Sampling bias** | When the sample you are drawing from is not representative of a larger population. Unless you get responses from every single person in a workshop, for instance, you will have a biased sample. For example, I might send out a workshop evaluation survey on a day when a third of the workshop attendees are at a conference, so are not able to respond. | Check whether responders had similar profile distributions to those who attended the workshop. Compare demographics (gender, domain, career stage, etc.) to help detect bias even in anonymous surveys. However, such information could be used to identify individuals in a smaller workshop. |
| **Social desirability bias** | A person responding to questions wishes to give a response that will make the interviewer think well of them. For example, I might feel uncomfortable answering the question "After this workshop on measuring impact, I feel confident about measuring the impact of my next event," if after the workshop I still didn't understand the topic. | Questions can emphasise the need for honesty and promise that although answer will be used and published, respondents will remain anonymous. For questions that ask about skill levels before and after a workshop (e.g., Rule 8), it is very important to indicate that it is OK if the respondent does not know how to do a skill. |

## Controlling for bias

To control for bias, consider which biases will most likely affect the results of your study and determine strategies to counteract those biases to the best of your ability. Be conscious of the fact that bias always exists and consider how it will affect your analysis. For example, use best practices in asking questions in survey research (Rule 5).

# Rule 5: Design your surveys well

As part of a wider range of social research methods (22), surveys are a key mechanism for evaluating workshops. They can form part of the information gathering before people attend (e.g., during registration), at the workshop (e.g., for unconferences (23)), after the workshop (e.g., as feedback forms), and much later after the workshop in follow up or impact surveys (see Rule 8).

## A note on quantitative vs. qualitative

Two types of survey questions can be asked: quantitative and qualitative. Quantitative questions are usually answered by many respondents and have definitive answers. They often use Likert scales, where respondents indicate how much they agree or disagree with a statement by choosing from set of fixed choices on a linear scale (e.g., strongly agree, agree, neither agree or disagree, disagree, strongly disagree) (24). Qualitative questions are more open-ended, and the answers can be probed using thematic analysis (25). Answer to qualitative questions allow you to gather information about the workshop and formulate hypotheses. They can even guide you as to which quantitative questions you could ask in the future, by helping you to identify the concepts your current questions are missing.

## Common pitfalls and how to avoid them

The questions asked and how they are constructed are an important part of survey design. Table 2 details what to watch out for when constructing survey questions to decrease bias and increase the clarity of what is being asked, thereby improving the quality of the results. The overall goal is to design straightforward questions that respondents can easily understand and answer. A common response by participants to any barrier to answering a question, from technical difficulties or confusingly worded questions, is to not complete the survey.

*Table 2: How to decrease bias and increase clarity in survey questions*

| Aspect | Explanation of issues | How to counter |
|---|---|---|
| **Compound question** | They are complex, overly wordy, and have multiple potential answers. An example of such a question: "*Would you prefer if this workshop were offered on Thursdays for two hours or Fridays for two hours or do you not care which day it's offered, but only that it's offered every week?"* These questions are hard to answer clearly and often only one portion of the question will be answered. Analysing the answers is thus potentially meaningless. | Deconstruct the compound question into separate questions. |
| **Leading question** | These questions guide the respondent towards a particular desired response. In combination with the social desirability bias, this is one of the easiest ways for survey research to become biased. For example: "*Given the number of people who have expressed an interest in weekend workshops, how interested are you in signing up for the workshop?"* | Remove any leading parts to the question: *"How interested are you in signing up for the workshop?"* |
| **Complex question** | These questions are a challenge for the respondent to follow and to accurately respond to. Similar to the compound question, it makes it hard both for the respondent to answer accurately and for the researcher to know what is being measured.<br>For instance: "*Imagine you are trying to teach a student who has never used the command line how to use it to do a pull request in GitHub. What are the ways you would teach that to someone from a different background than your own and how would you relate that to the teaching you would do of a loop in Python?"* | Pre-test your survey so that these types of questions can be highlighted and re-worded before your run the survey for real |
| **Multiple-choice question** | Multiple-choice questions that do not offer all of the possible answers included are naturally difficult to accurately respond to. For example, a question asking for a report of eye colour that does not include the respondent's eye colour in the possible answer choices cannot be answered. | Undertake qualitative work and/or pre-test your survey to find all of the possible answers to your multiple-choice questions.<br><br>Include an "other" response text box to capture other categories. Some manual coding and/or cleaning (26) will be needed to make use of the data. |
| | The order of the multiple-choice answers should be intuitive and have a flow. In some cases, it might make sense to randomise the choices to control for bias. In other situations where confusion could be caused (e.g., standard lists of ethnicity or domains), keeping a logical order is less confusing. Confusing those who fill in the | Check whether the answer choices should be randomised or kept in a logical/standard order |
| | survey is a sure way of decreasing response rates. | |
| **Wording choice** | It is rare that the use of absolutes such as “always” or “never” will help you write an effective survey question. Using an absolute in a survey question can mean that the response is not as useful because the respondent may have one instance that rules out an answer, e.g., “*Do you prefer workshops to always be run on Tuesdays?*” | In the majority of cases, remove or replace any absolute word(s) in questions. |
| | Keep answers comparable between respondents. For example, asking a respondent if they travelled “far” to attend the workshop could be subjective, with some people considering 10 miles to be far and others considering over 100 miles to be far.<br><br>It is equally important to manage value-laden words, such as ‘good’ and ‘bad’. However, value-laden and subjective questions can be useful for qualitative analysis of the workshop and can help to understand respondents’ perspectives, leading to future quantitative questions. | Define what you mean when asking about matters that are open to subjective opinion. E.g., rather than “far,” you could give a selection of distances. “Good” could be replaced with something more specific about your intent |
| | | such as “useful” or “enjoyable,” depending on what you are trying to measure. |
| **Open-ended question** | Not offering one open-ended question can cause you to lose out on information from attendees. | The question allows respondents to highlight anything positive or negative about the workshop that they would like to bring up. This can act as an additional safety net to catch issues with the survey that may have slipped through pretesting. |

# Rule 6: Ask about participants’ “confidence”

A common question that you can ask both at the start and end of a workshop is “How confident are you about <workshop topic>.” This question allows you to gauge the participants’ change in confidence and analyse whether the workshop changed the level of confidence about a particular subject, or technique, or ways of working together (for e.g. creating workshops).

It is possible that participants’ confidence might actually decrease as they realise that they know less than the other participants or they discover that there is much more to a particular field than they first realised. However, we only rarely found this to be the case in the Collaborations Workshop (10) series run by the Software Sustainability Institute (27). On average, confidence levels at registration compared to confidence levels in the post-course feedback showed an increase. If your participants report decreased confidence, look closer at the reasons why by mining the responses to open-ended questions or by following-up with the participants. You could include an open-ended question asking about the direction and cause of the change in confidence, if you suspect increased understanding may decrease confidence.

Asking about confidence has some limitations. For example, those answering the pre-course question about confidence level may not be the same people answering the post-course question, which prevents you from getting a true representation of the change in confidence level. You can mitigate this issue by asking participants to gauge their confidence level both pre- and post-course in the same post-course survey. You can then compare the average change across the same set of participants.

The term confidence can mean different things to different people; it might mean “how well you know the area” or “how well you can do something” or “how well you can explain something." This is ok, as what we are looking at is an individual's perceived change in confidence, whatever that means for them. If you are interested in just one of these possible interpretations, rephrase the question or add another question to ask about competence as well as confidence so that you can capture changes related to overall skills in an area, for example.

Asking about confidence is helpful if you want to know whether your workshop has made a difference for a particular field, area, or technique. If your goals require you to measure the change in your participants' skills, then Rule 7 (ask about specific skills) and Rule 9 (test specific skills) will be more important for you. These two rules are especially relevant measures for learning workshops.

# Rule 7: Ask about specific skills

As explained in Rule 1, all workshops should have objectives set. Objectives for learning workshops tend to be acquiring or mastering of specific skills or techniques. Objectives for exploring workshops tend to be knowledge, understanding, and an idea of where to look for more information or to find collaborators. Objectives for creating workshops tend to be learning a new skill, feeling like you have contributed towards a project, being able to do things differently, or finding future people to work with.

Although it is easy to ask questions around improvement in confidence (Rule 6), these questions are often too broad on their own. For deeper insights into the workshop's impact on its attendees, we need to craft more in-depth questions geared specifically to measuring the objectives (learning or otherwise) for those attending. You could ask people about their different levels of agreement for specific skills after the workshop using a Likert scale (e.g., strongly agree, agree, neutral, disagree, strongly disagree).

Examples of specific skill questions are:

- I understand the purpose of <a particular technique>
- I can describe the <process>
- I can apply the <technique> to my work
- I have a firm plan for how I am going to introduce what I have learned from this workshop into my work

It can be difficult to repeat these sorts of questions in surveys immediately after the initial post-workshop survey and in a six-month post-workshop follow-up survey, as they are specific enough for participants to have forgotten the details. However, using the "write to their future self" approach mentioned in Rule 8 can help to remind participants of their planned objectives before sending them a follow-up survey.

# Rule 8: Gather feedback before, during and after

A post-course feedback questionnaire is not the only way to measure the impact of a workshop. There are a number of other times in the process when asking participants questions can help to both run a more effective workshop and measure impact.

## Before

When running a workshop, it is important to collect demographic information of registrants, such as domain and career stage, to ensure that your audience is representative of the people you are targeting. This can be done at the pre-course stage. Other specific data to capture are their learning expectations, what they hope to discuss during the workshop, and what their (perceived) existing competencies are in the subject. This information is useful for structuring the workshop and adapting the content to align with the registrants' needs.

A good time to gather this pre-course information is at the point of registration. Participants are keen to attend the event and have a clear idea of why they are signing up and are thus likely to provide what is asked.

## During

During multi-day events, feedback of how the event is going can be collected at the end of each day and fed back to facilitators and organisers. This kind of ongoing feedback allows you to identify and respond to problems as they occur. You can also keep a running score of how well participants feel the event is meeting its objectives, or even see whether participants change their goals or what they feel the objectives of the event should be as the event goes on.

## At the end

At the end of a workshop, participants are normally bubbling with ideas, techniques they have learnt, things they want to change about their work upon their return, and which of the people they have met they might follow up with. However, normal life can sometimes take over, with pressing deadlines and the same old environment distracting the participants from carrying out their plans.

An excellent exercise is to ask participants to write to their future selves (28). In this technique, the positive change envisaged by those who attended the workshop is captured in written form at the moment they are most enthusiastic. An example of how to run this exercise is to ask the participants during the last session of the workshop to write a postcard to themselves as a reminder for future actions. You could ask the participants to write about how they want to use what they have learned, or how they would like to change some aspects of their current practise as a consequence of attending the workshop, or what their action plan is.

The postcards are then collected up by the organiser. Sometime after the workshop has closed, say two months to four months later, these postcards can be sent out to each of the participants, as a reminder of what they planned. There is something intriguing about physical postcards in the age of digital communication, which only adds to the impact of such practices.

While evaluation questionnaires can help measure the impact of workshops, this technique is a fun, but innovative way to extend the impact of the workshop beyond the time in which it was run.

## After

Post-course feedback should be collected soon after the workshop is completed, whilst things are still fresh in people's minds. Ideally, if the survey is hosted online, the participants can be given access to the link during the course, reminded in person on the last day, and then reminded again by email within two days, thereby maximising the chance of responses.

## Much After

Assessing the long-term impact and influence of your workshop on the individual can be difficult. To assess if your workshop has, or is, making a difference, send out a survey sometime after your workshop has completed (e.g., four to six months after the workshop). You can ask questions about what the participants learned at the workshop, how they have applied this knowledge to their work, and what impact the knowledge and network has had on their working life and practices. Another option is to conduct one-on-one interviews with participants. Although time consuming to conduct, those who are willing to talk can offer a lot of useful information and a much more nuanced view of impact than a survey.

Some workshops will involve a cohort of participants that remain linked together after your workshop. For instance, your workshop might form one of a series that the participants will all attend, or they might attend your workshop as part of their degree training as a group. The participants might even form themselves into a cohort that did not exist before the workshop, choosing to remain connected after the workshop through regular meet-ups. Such cohorts can make it easier to get feedback. For cohorts that are formed before your workshop, you can factor when they will meet when you plan the intervals at which you will get feedback. For example, you could arrange recorded interviews with a selected number of participants from the cohort during one of their scheduled meetings as another way of collecting feedback. The recordings can then be used to promote the workshop, maximise its impact, and provide evidence to funders of how people used what they learned.

# Rule 9: Harness gamification to test participants' skills

In Rule 7, we asked participants whether they felt that they had acquired certain skills. We can also test whether they have acquired these skills. One way to assess if people have learned a particular skill from your workshop is to assess them indirectly through an informal learning and assessment platform. Asking the participants to play a game alleviates the features of standardised testing environments that can cause anxiety. In games, learners encounter materials in new ways and have to apply their learning, not just repeat memorised details, and must rely on tacit knowledge. Games can show whether they have understood core concepts and knowledge areas. They can also highlight gaps and thus better focus the efforts of future workshops. For all its benefits, game-playing remains underused, although some examples can be pointed to as useful case studies.

Such complementary assessments fall into the category of “serious games” or “games with a purpose” (29), with an example being the Treasure Explorers (30). This tool combines different question types (multiple choice, tagging, and connecting ideas) as a way to help quantify people's understanding. The system was evaluated using games created to test understanding of logic and language, following the Brazilian National Educational Plan. People who use the system don’t feel like they are being formally tested. The system also has a social element, which shows a leaderboard (connected to players’ Facebook accounts) comparing how well players have done and allowing players to post their scores on social media. This competitive element again adds to what is termed “playful learning.”

Developing a game-playing assessment system from scratch can be time consuming, but there are toolkits that can help (31) (32), For a longer running workshop or a training series, it could form a worthwhile part of the evaluation method. Given the nature of such a system, it could even work with Rule 8: players (learners) can be sent a reminder to play, perhaps on a monthly basis, to keep the knowledge fresh in their minds and encourage them to use it in their day-to-day work (33). A useful output from a creating workshop might be to make such systems for education, assessment, or even to solve parts of computational pipelines in their domains (34).

# Rule 10: Measuring those who did not attend

It is easy to forget to measure the impact your workshop had on those who did not attend. In today’s social world, both organisers and participants have even more ways of sharing their content and what they have learnt “beyond the room.” Live recordings, blog posts, tweets, Instagram photographs, and standalone reports are all ways to allow your workshop to keep reaching new audiences beyond the workshop date. Encourage your attendees to share their experiences during and after your sessions. Twitter is currently a favoured platform for such event amplification (35). If you want to encourage event amplification, use a uniform hashtag across promotional material and, resources permitting, have one of your organisers actively contribute to and monitor the conversation during the workshop. This will increase the workshop's impact and your interaction with those who are not in physical attendance. Take the time to produce a report after the event yourself and share it in a venue that gives it a permanent Digital Object Identifier (DOI) (36) so it is easier for you to track citations.

Wherever your workshop information is shared, in whatever format, you have the chance to measure the impact of that information beyond the workshop's original remit. Effort is needed to track this impact through citations and other measures of views and use, for example by using systems such as Altmetric (37), Google Analytics (38), YouTube, figshare (39), SlideShare (40), and Twitter Analytics (41). This effort will help you to show the impact from your events and should form part of your overall measurement. These statistics should be tracked regularly, perhaps every six months or annually.

Another metric of impact beyond the room is whether participants talk about their experience positively with friends and colleagues. If you are running a workshop series, you could track recommendations by asking participants how they found out about the workshop. Referrals from previous participants is a good sign that you are doing something right.

# Conclusion

It is clear that you need to plan (Rule 1 and 2), use your knowledge & skills (Rule 3, 4 and 5) and apply techniques (Rules 6-10) to be able to measure the impact of workshops (those focused on exploring, learning or creating or a mix). Ultimately, it is worth understanding why we want to measure impact in the first place and balance this with the amount of time required to organise the workshop and time we want to put into evaluating the workshop. With good measurements, we can convince funders to maintain and support the work that we do, encourage people to attend our workshops and feel satisfied that the work that we are doing with our workshops is worthwhile and making a positive difference.